\documentstyle[twocolumn,aps,epsfig,floats]{revtex}
\begin{document}
\draft

\title{A Fluctuation-Dissipation Process without Time Scale}
\author{Mario
Annunziato$^1$, Paolo Grigolini$^{1,2,3}$ and Juri Riccardi$^{4}$}
\address{$^{1}$Dipartimento di Fisica dell'Universit\`{a} di Pisa,
Piazza
Torricelli 2, 56127 Pisa, Italy }
\address{$^{2}$Center for Nonlinear Science, University of North
Texas, P.O. Box 305370, Denton, Texas 76203 }
\address{$^{3}$Istituto di Biofisica del Consiglio Nazionale delle
Ricerche, Via San Lorenzo 26, 56127 Pisa, Italy }
\address{$^{4}$ENEL SpA Struttura Ricerca-Area Generazione Via A. Pisano 120, 56100,
Pisa, Italy}
\date{\today}
\maketitle

\begin{abstract}
We study the influence of a dissipation process on
 diffusion dynamics triggered by fluctuations with long-range
correlations. We make the assumption that the perturbation process
involved is of the same kind as those recently studied numerically
and theoretically, with a good agreement between theory and
numerical treatment. As a result of this assumption the
equilibrium distribution departs from the ordinary canonical
distribution. The distribution tails are truncated, the
distribution border is signalled by sharp peaks and, in the weak
dissipation limit, the
 central distribution body becomes identical to a truncated
 L\'{e}vy distribution.
\end{abstract}


\section{introduction}

The derivation of thermodynamics from dynamics is still an open
field of investigation \cite{HHPP87,PP96}. Hereby
 we focus on a related but seemingly less
  ambitious purpose, the derivation of fluctuation-dissipation
processes
  from deterministic dynamics \cite{J95,BMWG95}.
  It has been recently pointed out\cite{BMWG95} that
   a genuinely dynamic derivation of Brownian motion
   would be essentially equivalent to a mechanical
   foundation of thermodynamics, thereby implying that also
   this avenue might be fraught by strong conceptual difficulties.
    It has been remarked\cite{BMWG95} that the dynamic foundation of
    Brownian motion, as described by an ordinary Fokker-Planck
equation,
     imply fluctuations with a finite correlation time $\tau$,
     namely, rests on the existence of a finite \emph{microscopic
     time}, or,
     equivalently, on the microscopic foundation of the linear
response
     theory\cite{BMG96}. However, the resulting transport equation
can be
     identified with a \emph{bona fide} Fokker-Planck equation only
if\cite{F77}
     the corresponding relaxation process is exactly, not
approximately,
      exponential: a property in harsh conflict with both quantum
      \cite{FGR77}
       and classical\cite{L83} dynamics. This is the main reason why
the problem
        of the dynamic foundation of the ordinary Fokker-Planck
equation is
        not yet settled and further efforts must be made not
excluding the
        possibility of either non-Newtonian effects \cite{PP96} or
spontaneous
        fluctuations \cite{TGV95}, both implying a kind of generalization
of
        ordinary classical and quantum mechanics.
    Here we reverse the perspective and rather than
    imposing the Markovian approximation, incompatible
     with the deterministic nature of the system under study,
     we discuss the consequence of explicitly rejecting the
     requirement of a finite microscopic time scale. To conduct
     this discussion, we adopt Occam's principle, namely we study
      the simplest dynamical system with the essential features
       necessary to produce a fluctuation-dissipation process
        without using the requirement of a finite microscopic
        time scale.

        Let us consider the Liouville-like equation
        \begin{eqnarray}
       \frac {\partial}{\partial t} \rho_{T}(x,\xi,{\bf w},t)
       &=& \hat{L} \rho_{T}(x,\xi,{\bf w},t)
        \nonumber \\
       &\equiv&[ - \xi \frac{\partial}{\partial x} +
       \hat{\Gamma}(-\Delta^{2} x)]\, \rho_{T}(x,\xi,{\bf w},t).
       \label{thewholeuniverse}
       \end{eqnarray}
      This means that we study for simplicity a one-dimensional case.
      The one-dimensional variable of interest $x$  undergoes the
      influence of a ``fluctuation", called $\xi $. The dynamics of $\xi$
      is driven by the operator  $\hat{\Gamma}$ which expresses concisely the
      action that a set of variables $\mathbf{w}$  can exert on $\xi$ so as to
      render disordered its time evolution. Thus, in principle,
      the "stochastic" dynamics of $x$ can be either provoked by
      nonlinearity or by the large number of degrees of freedom.
      The unperturbed fluctuations  $\xi$ are the source of diffusion
      of the variable $x$. To undergo also dissipation, the second key
      ingredient of a fluctuation-dissipation process, the variable $x$
      must also exert a feedback on the dynamics of $\xi$. This important
      property is expressed by the dependence of $\hat{\Gamma}$ on
      $\Delta^{2}$,  left unspecified. As we shall see, we shall assume a
      linear departure from the unperturbed condition $\hat{\Gamma}_{0}$
      given by $\hat{\Gamma}(-\Delta^{2} x) = \hat{\Gamma}_{0}
      -\Delta^{2}x\hat{\Gamma}_{1}$. The operator $\hat{\Gamma}_{1}$
      drives the bath response to an external perturbation\cite{BMWG95}.
       The parameter $-\Delta^{2}\,x$ denotes the strength of the feedback
        and the minus
       sign alludes to the reaction nature of the effect.
    Pursuing our program inspired to Occam's criterion, we are forced
    to assume the variable $\xi$ to be dichotomous. It would be surprising
    if $\xi$,
    and so the microscopic statistics, would be Gaussian. In a sense,
there
    would be no problem to solve at all. Therefore we must adopt a non
    Gaussian statistics. Thus, we fix the statistics of $\xi$ to be
dichotomous,
    since dichotomous statistics seem to be the simplest example of
    non-Gaussian statistics.

    We plan to prove that a bath, described by the unperturbed operator
    $\hat{\Gamma}_{0}$ with  a diverging
    correlation time, $\tau = \infty$, yields
 a form of equilibrium
    strongly departing from the ordinary canonical prescription. The
    proof is organized as follows. In Section II we derive a generalized
    master equation. We make the basic assumption that the memory
kernel of
    this master equation only depends on the unperturbed bath dynamics.
    Using the additional assumption that the bath response is of the
    same kind as that studied in earlier publications, we derive the
    central theoretical result of this paper. In Section III we discuss
    the error associated with the basic assumption of Section II. In
    Section IV we use the central theoretical result of this paper to
    predict the resulting, non-canonical, equilibrium. In Section V we
    check numerically the effect of the error discussed in Section II.
    Finally, in Section VI we make a balance of the results obtained in
    this paper.

\section{the projection method}
In the case of no feedback, namely when $\Delta = 0$, it is shown \cite{AGW96} 
that an immediate benefit of the dichotomous choice is that the 
projection operator method \cite{Z61} allows us to express the dynamics of 
the variable of interest, x, in terms of an exact and simple 
diffusion-like equation of motion.
When the feedback is included, unfortunately, the
    projection method  does not produce a simple
    equation of motion, and delicate assumptions must be made
if we want to keep the elegance and simplicity of the earlier
treatment. The main purpose
     of this section is to discuss these delicate assumptions.
  The adoption of the
     projection method \cite{Z61} yields for the part of interest
$\hat{P}\rho_{T}$
      of the total
     distribution the following equation:
      \begin{eqnarray}
       \frac {\partial}{\partial t} \hat {P}\rho_{T}(x,&&\xi,{\bf w},t)
       = \int_{0}^{t}\Big\{\hat{P}[-\xi \frac{\partial}{\partial x}]
       \exp[\hat{Q}\hat{L}\hat{Q}(t-t^{\prime})]\cdot\nonumber \\
       &&  \hat{Q}
       [-\xi \frac{\partial}{\partial x} + \Gamma(-\Delta^{2}x)]
       \hat{P}\rho_{T}(x,\xi,{\bf w},t^\prime)\Big\}\,dt^{\prime}.
       \label{generalizedmasterequation}
       \end{eqnarray}
       As usual,
      we assume the bath to have an equilibrium distribution satisfying
      the condition:
      \begin{equation}
      \hat{\Gamma}_{0}\rho_{eq}(\xi, {\bf w}) = 0.
      \label{equilibriumdistribution}
      \end{equation}
      This dictates the choice of the projection operator $\hat{P}$:
       \begin{equation}
       \hat{P} \rho_{T}(x,\xi,{\bf w}; t)
      = \sigma(x,t) \rho_{eq}(\xi, {\bf w}) ,
      \label{projectionoperator}
      \end{equation}
      where $\sigma(x,t)$ is obtained tracing the total distribution
      $\rho_{T}(x,\xi,{\bf w},t)$ over the irrelevant degrees of freedom
      $\xi$ and ${\bf w}$.
 Note that Eq.(\ref{generalizedmasterequation}) is an exact
equation provided that the initial
       condition is given by:
       \begin{equation}
      \rho_{T}(x,\xi,{\bf w}; 0)
      = \sigma(x;0) \rho_{eq}(\xi, {\bf w}).
      \label{factorizedform}
      \end{equation}
 For simplicity  we assume that:

  \begin{equation}
      \hat{\Gamma}(K) = \hat{\Gamma}_{0} + K \hat{\Gamma}_{1}.
      \label{linearassumption}
      \end{equation}

      We  carry out our calculations setting the condition:
      $K=-\Delta^2\,x$. This is equivalent to assuming a form of
      linear response to external perturbation in agreement with
      Refs. \cite{TFWG94,barkai1,barkai2,klaftersubdiffusional}.
       We make now the assumption that, in spite
      of $\Delta \neq0$, the exponential operator $\exp(\hat{Q}\,\hat{L}\hat{Q}(t-t')$
       appearing in Eq. (\ref{generalizedmasterequation}) only depends on the
       unperturbed
       operator $\hat{\Gamma}_0$, a property that, as earlier pointed out, is valid only
       in the free diffusion case.  We also use Eq.(\ref{equilibriumdistribution}).
        Under all these approximations, we are allowed to
  rewrite Eq.(\ref{generalizedmasterequation}) as
\begin{equation}
       \frac {\partial}{\partial t} \hat {P}\rho_{T}(x,\xi,{\bf w},t)
       = A(\rho_{T}(x,\xi,{\bf w},t)) + B(\rho_{T}(x,\xi,{\bf w},t))
       \label{addition},
       \end{equation}
where
\begin{eqnarray}
A(\rho_{T}(x,\xi,{\bf w},t)) \equiv     \frac
{\partial^{2}}{\partial x^{2}} \int_{0}^{t}dt^{\prime}\Big\{ \hat
{P}[\,\xi\,]  \exp[\hat{\Gamma}_{0}(t -
t^{\prime})]\cdot\nonumber\\
 \hat{Q}[\,\xi\,] \hat{P} \rho_{T}(x,\xi,{\bf
 w},t^{\prime})\Big\}
       \label{A}
       \end{eqnarray}
and

\begin{eqnarray}
B(\rho_{T}(x,\xi,{\bf w},t)) \equiv    \Delta^{2} \frac
{\partial}{\partial x} x \int_{0}^{t}dt^{\prime}&& \Big\{\hat
{P}[\,\xi\,]
 \exp[\hat{\Gamma}_{0}(t - t^{\prime})]\cdot\nonumber \\&& \hat{Q}[\,\hat{\Gamma}_1\,] \hat{P} \rho_{T}(x,\xi,{\bf
 w},t^{\prime})\Big\}.\nonumber \\
       \label{B}
       \end{eqnarray}
In conclusion to derive this result we have used a major
assumption that will be referred to as \emph{assumption (i)}.This
assumption can be expressed as follows:

\emph{assumption(i)}. We assume that the exponential operator
appearing on the \emph{r.h.s} of
Eq.(\ref{generalizedmasterequation}) only depends on the
unperturbed bath dynamics.

In the case where
no feedback process is considered\cite{AGW96} there is no error
associated with this assumption, since this is shown to be an exact
consequence of the dichotomous nature of the variable $\xi$. It is not
so in the more general case of this paper. We shall devote Section III
and the numerical treatment of Section V to assessing the consequences of
the error associated with this basic assumption.
    Adopting the formalism of the response theory\cite{BMWG95} we  rewrite
    Eq.(\ref{addition}) in the form:
          \begin{eqnarray}
       \frac {\partial}{\partial t}\sigma(x,t)&
       =& <\xi^{2}>_{eq} \int_{0}^{t}dt^{\prime} \,\Phi_{\xi}(t-t^{\prime})
       \frac{\partial^{2}}{\partial x^{2}}
       \sigma(x,t^{\prime})\nonumber \\
       &+& \Delta^{2}<\xi\,\Gamma_{1}>_{eq}\int_{0}^{t}dt^\prime\,C(t-t^{\prime})
       \frac{\partial}{\partial x}x \sigma(x,t^{\prime}),
        \label{fundamentalequation}
       \end{eqnarray}
where
\begin{equation}
      \Phi_{\xi}(t) \equiv \frac{<\xi\, \xi(t)>_{eq}} {<\xi^{2}>_{eq}}
      \label{correlationfunction}
      \end{equation}
and
      \begin{equation}
      C(t) \equiv \frac{<\xi \,\exp(\hat{\Gamma}_{0}\,t)\,\hat{\Gamma}_{1}>_{eq}}
      {<\xi \,\hat{\Gamma}_{1}>_{eq}}.
      \label{correlationfriction}
      \end{equation}
    This result has been obtained by evaluating the diffusion term at
    the zero-th order in the feedback interaction, and considering,
    in agreement with the linear response criterion\cite{BMWG95}, the
first
    non-vanishing contribution, proportional to the friction.
     It is worth remarking that the correlation function of
    Eq.(\ref{correlationfunction}) affords the most convenient way of
    defining the microscopic time $\tau$ mentioned in Section 1. This
is given
    by:
    \begin{equation}
    \tau \equiv \int_{0}^{\infty}\Phi_{\xi}(t)dt.
    \label{correlationtime}
    \end{equation}

    We now have recourse to the second approximation on which our crucial
    theoretical results rests.  This assumption will be referred to as
\emph{assumption (ii)} and can be expressed as follows:

\emph{assumption(ii)}. We assume that the function $C(t)$ has a finite
time scale.

This assumption is dictated by the theoretical and numerical
conclusion of the earlier work of
Refs.\cite{TFWG94,barkai1,barkai2}. This assumption cannot be
mislead as a property of ordinary statistical mechanics. Actually,
this assumption means a deviation from ordinary statistical
mechanics, which, as shown in Ref.\cite{BMWG95}, would imply
\begin{equation}
C(t) = \Phi_{\xi}(t).
\label{ordinarystatisticalmechanics}
\end{equation}
In the case where the correlation function $\Phi_{\xi}(t)$ is not
integrable and the correlation time of Eq.(\ref{correlationtime})
diverges, the condition of Eq.(\ref{ordinarystatisticalmechanics})
would imply a field of finite intensity to produce a current of
infinite intensity\cite{TFWG94}. The numerical calculations show
that this striking physical condition is not
realized\cite{TFWG94}, thereby implying a violation of Eq.
(\ref{ordinarystatisticalmechanics}). This violation, in turn, is
due to the fact that the function $C(t)$ has a finite time scale
even when the function $\Phi_{\xi}(t)$ does not.
    Note that under the assumption that the function $C(t)$ has a finite
    time scale it is possible to define
    \begin{equation}
    \gamma \equiv \Delta^{2}<\xi \,\Gamma_{1}>_{eq} \int_{0}^{\infty}
    dt^{\prime}C(t^{\prime}).
    \label{friction}
    \end{equation}
    In conclusion, we obtain the following important equation:

 \begin{eqnarray}
       \frac {\partial}{\partial t} \sigma(x,t)
       = <\xi^{2}>_{eq} \int_{0}^{t}dt^{\prime} \Phi_{\xi}(t-t^{\prime})
       \frac{\partial^{2}}{\partial x^{2}}
       \sigma(x,t^\prime)\nonumber \\
       + \gamma \frac{\partial} {\partial x}x \sigma(x,t).
        \label{fundamentalequationwithordinaryfriction}
       \end{eqnarray}
       This result rests on both assumption (i) and (ii). However, it is
       evident that special attention must be devoted to the first
       assumption.  In a sense the validity of assumption (ii) has already
       been assessed by the theoretical and numerical work of
       Refs.\cite{TFWG94,barkai1,barkai2}. The validity of assumption
       (i), on the contrary, requires further discussion. This will be
       done in Sections III and V.

Before ending this Section, we want to remark that in the special
    case where the condition of Eq.(\ref{ordinarystatisticalmechanics})
    applies, the important result of Eq.(\ref{fundamentalequation})
    becomes very similar
    to the Fokker-Planck type of equation
    recently found by the authors
     of Ref.\cite{klaftersubdiffusional}.
     These authors pointed out that an equation of this kind
     shows that anomalous diffusion can be compatible with
     Boltzmann statistics. We note that this conclusion does not apply
     to the case of super-diffusion under study in this paper, because, as
     we have seen, Eq.(\ref{ordinarystatisticalmechanics}) does not apply.
     In the subdiffusional case studied in
     Ref.\cite{klaftersubdiffusional}, however, there are no compelling
     reasons leading to the breakdown of
     Eq.(\ref{ordinarystatisticalmechanics}) thereby leaving open the
     possibility that the subdiffusional condition is compatible with 
     Boltzmann statistics.

    \section{no interference between free fluctuation and
    dissipation:time evolution}
    To a first sight, one might be led to think that
    Eq.(\ref{fundamentalequationwithordinaryfriction}) is equivalent to
    the Langevin-like equation:
    \begin{equation}
    \dot{x}(t) = - \gamma\, x(t) + \xi(t),
    \label{langevinequation}
    \end{equation}
    supplemented, of course, by the set of equations necessary to
    determine the time evolution of the dichotomous variable $\xi(t)$.
   In this Section we show that
    Eq.(\ref{fundamentalequationwithordinaryfriction}) is not identical
to the
    equation of motion for $\sigma(x,t)$ generated by
    Eq.(\ref{langevinequation}). This will help us to estimate the error
    affecting the main prediction of this paper about the condition of
    equilibrium established by the feedback on the generator of
    fluctuation without time scale.
    \subsection{Second moment time evolution}
    In Section IV we shall point out that Eq.(\ref{langevinequation})
implies
    that throughout system's time evolution the trajectory $x(t)$
    departing from the initial condition $x(0) = 0$ never leaves the
    interval $[- W/\gamma, W/\gamma]$. This property means that the second
    moment of the distribution is kept finite at all time and can never
    exceed the maximum value $(W/\gamma)^{2}$. Here we show that, on the
    contrary, the second moment of the distribution driven by
    Eq.(\ref{fundamentalequationwithordinaryfriction}) diverges for $t
    \rightarrow \infty$.

    Using Eq.(\ref{langevinequation}) we get:
    \begin{eqnarray}
    \frac{\partial}{\partial t}<x^{2}(t)>
    = \gamma \int_{\infty}^{-\infty} dx\, x^{2} \frac {\partial}{\partial
    x} [x \sigma(x,t)]\nonumber \\
    + <\xi^{2}> \int_{\infty}^{-\infty} dx\, x^{2} \int_{0}^{t} dt^{\prime}
    \Phi_{\xi}(t-t^{\prime}) \frac{\partial^{2}}{\partial x^{2}}
\sigma(x,t^\prime).
    \label{secondmomentfirstderivative}
    \end{eqnarray}
    Using the method of integration by parts it is shown that
    Eq.(\ref{secondmomentfirstderivative}) yields:
    \begin{equation}
        \frac{\partial}{\partial t}<x^{2}(t)>
        = - 2 \gamma <x^{2}(t)> + 2 <\xi^{2}> \int_{0}^{t}
        \Phi(t^{\prime})dt^{\prime}  .
        \label{secondmomenttimeevolution}
        \end{equation}
        Note that the first term on the \emph{r.h.s.} of
        Eq.(\ref{secondmomenttimeevolution}) can be derived from
        the first term on the \emph{r.h.s.} of
        Eq.(\ref{secondmomentfirstderivative}) via integration by
parts,
        provided that the decay of the function $\sigma(x,t)$ for
        $|x| \rightarrow \infty$ is faster than $1/|x|^{3}$. This means
        that the distribution $\sigma(x,t)$ cannot be a L\'{e}vy
process
        at any finite time $t > 0$. We know that at
        $\gamma = 0$ the diffusing distribution is in fact
        a L\'{e}vy process with ballistic peaks signalling the
presence of a
        propagation front\cite{AGW96}, thereby ensuring the
validity of
        the method of integration by parts. It is plausible to
assume that
        the action of a dissipation process makes the distribution
	spreading less intense, so favoring rather than opposing the method
	of integration by parts.

        The solution of Eq.(\ref{secondmomenttimeevolution}) is
given by:

        \begin{equation}
        <x^{2}(t)> = \frac{<\xi^{2}>}{\gamma}
        \int_{0}^{t}\Phi_{\xi}(t-t^{\prime})[ 1 - \exp(-2\,\gamma\,
        t^{\prime}) ] dt^{\prime}.
        \label{divergentsecondmoment}
        \end{equation}
        Let us adopt for the correlation function $\Phi_{\xi}(t)$ the
        choice:
        \begin{equation}\label{psidit}
        \Phi_{\xi}(t) =\frac{ (\beta\,T)^\beta}{(\beta\,T + t)^\beta}.
        \label{explicitform}
        \end{equation}
        where $T$ is the mean waiting time in a state of the
        velocity. In fact, as a consequence of the one-dimensional
         assumption we are allowed  to use the relation \cite{AGW96}:
        \begin{eqnarray}\label{permanencetime}
        \psi(t)=T\,\frac{d^2}{dt^2}\,\Phi_\xi(t)=\frac{(\beta\,T)^{\beta+1}\,
        (\beta+1)}{(\beta\,T+t)^{2+\beta}},
        \end{eqnarray}
        where $\psi(t)$ is the distribution density of sojourn
        times.
 	Plugging the analytical form of Eq. \ref{psidit} into the 
	{\emph r.h.s.} of Eq. \ref{divergentsecondmoment} and making a 
	time asymptotic analysis, we get:
        \begin{equation}
        \lim_{t \rightarrow \infty}<x^{2}(t)> \approx t^{1 - \beta}
, \,\,\,\gamma > 0
        \label{finitegamma}
        \end{equation}
        and
        \begin{equation}
        \lim_{\gamma \rightarrow 0 } <x^{2}(t)> \approx t^{2 -
\beta} ,\,\,\,t >> 1.
        \label{vanishinggamma}
        \end{equation}

        \subsection{Exact equation of motion for $\sigma(x,t)$}
        We note that the  use of the same  projection method as that
        applied in Section II to the dynamic system described by
        Eq.(\ref{langevinequation}) yields
        \begin{eqnarray}
        \frac{\partial}{\partial t} \sigma(x,t)
        = \gamma \frac{\partial} {\partial x} x \sigma(&x&,t)
        + \frac{\partial}{\partial x}
        \int_{0}^{t}dt^{\prime}\Big\{ \Phi_{\xi}(
        t-t^{\prime})\cdot\nonumber \\
        &&\exp[\gamma \frac{\partial}{\partial x} x(t - t^{\prime})]
        \frac{\partial}{\partial x} \sigma(x,t^{\prime})\Big\} .
        \label{exact}
        \end{eqnarray}
        We immediately see that the same approximation as that
applied to
        Eq.(\ref{generalizedmasterequation}), namely the
approximation of
        neglecting the influence of the feedback on the memory
kernel,
        makes Eq. (\ref{exact}) identical to
        Eq.(\ref{fundamentalequationwithordinaryfriction}).
Consequently,
        the numerical treatment of Eq.(\ref{langevinequation}) is
expected
        to depart from the prediction of
            Eq.(\ref{fundamentalequationwithordinaryfriction})
and   the amount of this departure can be used as a way to
            establish the  error caused by assumption (i) in
the derivation of
            Eq.(\ref{fundamentalequationwithordinaryfriction}),
which is the
            central result of this paper.

            Eq.(\ref{exact}) can be used to derive an
analytical expression for
            the second moment time evolution. The Taylor series
expansion of
            the exponential operator on the \emph{r.h.s.} of
            Eq.(\ref{exact}) and the use of integration by
parts yield:
            \begin{eqnarray}
\frac{\partial\,}{\partial t}<x^2(t)>+2 \gamma<x^2(t)> &=&
2<\xi^2>\cdot \nonumber \\
&&\int_{0}^{t}\Phi_{\xi}(t')\exp(-\gamma\,t')dt',\nonumber\\
\label{towardssecondmoment}
\end{eqnarray}
which, in turn, yields the following time evolution:
\begin{eqnarray}
<x^2(t)>=2<\xi^2> \exp(-2\,\gamma\, t)\int_{0}^{t}
\exp(2\,\gamma\, t^\prime)\cdot\nonumber \\
\int_{0}^{t^\prime}\Phi_{\xi}(\tau)\exp(-\gamma\,\tau)d\tau\,
dt^\prime . 
\label{newsecondmomenttimeevolution}
\end{eqnarray}
It is worth remarking that the general expression for the
asymptotic value of the second moment is:
\begin{equation}
<x^2(\infty)>=\frac{<\xi^2>}{\gamma}\int_{0}^{\infty}\Phi_\xi(t')\,\exp(-\gamma\,t')dt'\mbox{ .}
\end{equation}

 We see that the
asymptotic value for the second moment, as it must be, is
finite and in the special case of Eq.(\ref{explicitform}) the
analytical expression for the second moment at $ t =\infty$ is:
\begin{equation}
<x^2(\infty)>= <\xi^2>(\beta\,T)^{\beta}\exp(\gamma\,\beta\, T)
\frac {\Gamma (1-\beta, \gamma\,\beta\,T )}{\gamma^{2 -\beta}} .
\label{finiteasymptoticsecondmoment}
\end{equation}
In conclusion, we see that assumption (i) produces the seemingly
unacceptable effect of making the asymptotic second moment
diverge, whereas the exact equation of motion yields a second
moment which is always finite and gets at equilibrium the finite
value predicted by Eq.(\ref{finiteasymptoticsecondmoment}). The
discussion of Sections IV and V will explain in which sense the
error associated with assumption (i) does not invalidate our main
conclusion that the final equilibrium distribution is of L\'{e}vy
kind.

            \section{equilibrium properties}

            The fact that the second moment does not converge
to a finite value
            is a consequence of the central approximation yielding
            Eq.(\ref{fundamentalequationwithordinaryfriction}).
This does not
            conflict with the possibility that for $t
\rightarrow \infty$ the
            distribution approaches asymptotically a time
independent shape. Using the recent results of Refs \cite{AGW96}
and \cite{bologna} it is shown that the Fourier transform of
Eq.(\ref{fundamentalequationwithordinaryfriction}) obey the
following time evolution equation:
\begin{equation}
      \frac{\partial}{\partial t} \hat{\sigma}(k,t)
      = - b|k|^{\alpha}\hat{\sigma}(k,t)
      - \gamma k\frac{\partial}{\partial k} \hat{\sigma}(k,t),
      \label{almostfinal}
      \end{equation}
      where $\alpha=1+\beta$ and  $b$ is a positive constant (see Eq.(21) of
      Ref\cite{bologna}).
            In particular,  the paper of Ref \cite{bologna}  shows that
            in the asymptotic limit the contribution to free
       diffusion becomes indistinguishable from  that originally introduced
        by Seshadri and West\cite{seshadri}.
      Equation  (\ref{almostfinal}) yields the following equilibrium distribution:
      \begin{equation}
      \hat{\sigma}(k,\infty) = \exp\Big(- \frac{b}{\alpha\,\gamma} |k|^{\alpha}\Big),
      \label{westseshadri}
      \end{equation}
        which, in turns, according to\cite{west}, coincides with the
       equilibrium distribution corresponding to the equation of motion:
\begin{equation}
\frac{d}{dt} x(t) = -\gamma x(t) + \eta(t),
\label{stochasticequation}
\end{equation}
where $\eta(t)$ is an uncorrelated noise, with probability
distribution $p(\eta)$, obeying L\'{e}vy statistics,
and thus defined in the Fourier space by:
\begin{equation}
p(k) = \int d\eta \exp(- i k\eta) p(\eta) = \exp(-b\,|k|^{\mu}),
\label{levystatistics}
\end{equation}
where $0<\mu <2$.

It must be pointed out that Eq.(\ref{stochasticequation}) does not
coincide with Eq.(\ref{langevinequation}). In the case of very weak
friction they do in the sense that will be illustrated in Section V.

\section{numerical results}

\begin{figure}[t]
\begin{picture}(400,215)(-15,0)
\epsfxsize=9truecm
\epsfysize=7truecm 
\put(-25,15){\epsfbox{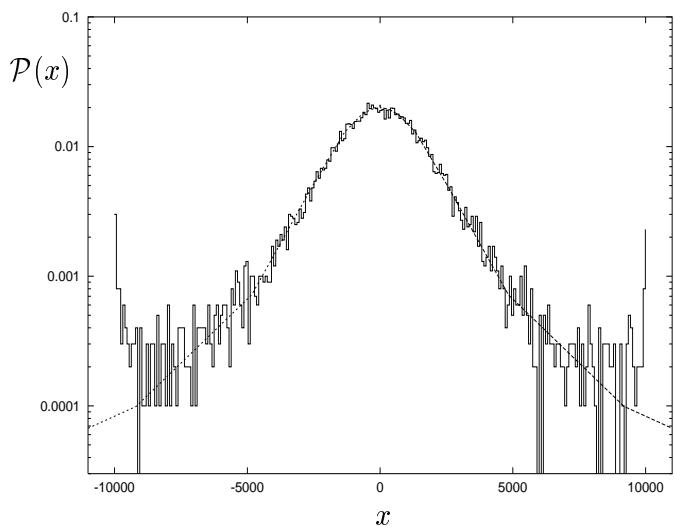}}
\end{picture}
\caption{Equilibrium distribution ${\cal P}(x) \equiv \sigma(x,\infty)$
as a function of $x$.
The L\'evy distribution obtained with the anti-Fourier
transform of Eq. (\ref{westseshadri}) (dashed line)
fits very well the distribution middle. The system 
parameters used are, $\gamma=10^{-4}$, $T=50$, 
$\beta=0.6$. The numerical conditions are: number of trajectories $=10^4$,
observation time $= 2\cdot 10^5$, bin interval $=80$ and $b = 8.4\cdot 10^4$. }
\end{figure}

The numerical results of this Section are based on the numerical
treatment of Eq.(\ref{langevinequation}), and consequently on the
numerical implementation of
\begin{equation}
     x(t)=\int_0^t  \exp[-\gamma(t-t^\prime)]\xi(t^\prime)\, dt^\prime
     + x(0)\,\exp(-\gamma t)\mbox{ .}
\label{langevinintegral}
\end{equation}
The fact that the variable $\xi$ is dichotomous with the correlation
function of Eq.(\ref{explicitform}) naturally leads us to adopt
the same numerical approach as that used in
Refs.\cite{AGW96,marcoluigi}. This means that two random number
generators are used. The first results in random number
homogeneously distributed in the interval $[0,1]$. With a proper
non-linear deformation this is made equivalent to a random
generation of  waiting times with the distribution of
Eq.(\ref{psidit}). This is the way we adopt to build numerically
the time evolution of $x(t)$. We also set an initial condition
fitting the crucial condition of Eq.(\ref{factorizedform}) and we
make the trajectory run for times larger than $20/\gamma$. We
evaluate the same way $10^{4}$ trajectories, then these
trajectories are recorded in a bin.

  In Fig 1 we see a sample of the resulting equilibrium 
distribution with $\beta=0.6$, $T=50$, $W=1$,
$\gamma=10^{-4}$, spanning from $ x = -W/\gamma$ to $x =
W/\gamma$. Fig. 1 is a crystal clear illustration of what we mean
by statistics of L\'{e}vy kind. We see that the equilibrium
distribution is truncated and that at the borders two sharp peaks
emerge. These sharp peaks are an equilibrium reflection of the
peaks already revealed by the numerical treatment of free
diffusion\cite{AGW96,ZK93}. However, the distribution enclosed by
these peaks is shown to fit very well the L\'{e}vy distribution
predicted by the theory  of Section IV.
\begin{figure}[t]
\begin{picture}(400,215)(-20,0)
\epsfxsize=9truecm 
\epsfysize=7truecm
\put(-35,15){\epsfbox{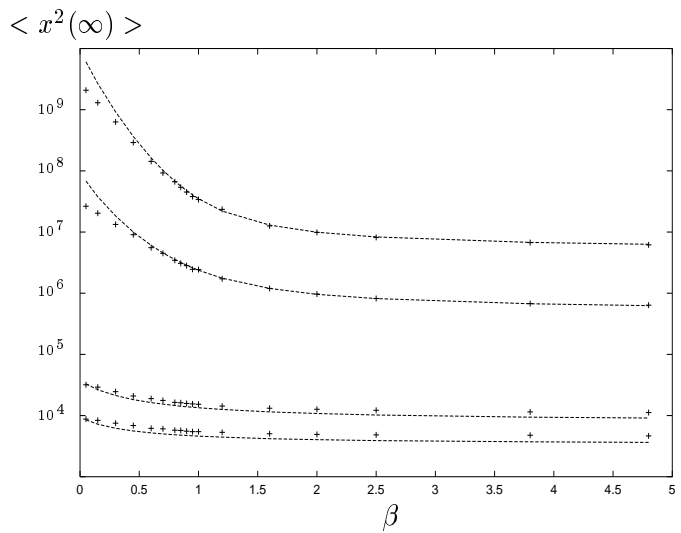}}
\end{picture}
\caption{Comparison between computed and theoretical 
distribution variance. The theoretical  prediction used is
Eq. (\ref{finiteasymptoticsecondmoment}). The crosses indicate
the numerical result and the lines the theoretical prediction.
Each curve has been obtained keeping $\gamma$ constant.
From the bottom to the top curve the values of $\gamma$ are: $10^{-2}$,
$5 \cdot 10^{-3}$, $10^{-4}$, $10^{-5}$.}
\end{figure}

  Fig. 2 is devoted to the
comparison between the theoretical prediction of
Eq.(\ref{finiteasymptoticsecondmoment}) and the result of our
numerical treatment. The agreement between theory and numerical
calculation is extremely good, not only in the Gaussian regime
$\beta >1$, but also in the regime $\beta <1$ up to $\beta \simeq 0.5$ . 
Significant discrepancies between theory and numerical
results can be found in the region close to $\beta= 0$, probably 
as a consequence
of the fact that with a finite number of trajectories the peaks,
which are expected to give significant contribution to the second
moment, cannot be satisfactorily reproduced.

In conclusion, these numerical results prove that we are in a full
control of the error caused by approximation (i). The markedly
L\'{e}vy character of the equilibrium distribution in the sense
here illustrated with the help of Fig. 1 supports our theoretical
prediction: The main feedback effect is the emergence of the non-canonical
statistics of Ref. \cite{west}.

\section{concluding remarks}
Which is then
 the interest of our results? We think that
 the interest of them lies on this: This paper
  forces us to change the conventional perspective
   concerning the microscopic foundation
   of the canonical statistical behavior.
    Some years ago, the findings of
    Zhu and Robinson\cite{ZR89} have been
    criticized by Keirstad and Wilson\cite{KW90}
    with arguments which are a nice example
    of the conventional wisdom. Let us see why.
    Zhu and Robinson \cite{ZR89} had detected
    significant deviations from the canonical
     Maxwell velocity distribution, in a
     physical condition characterized by
     a system of interest very fast compared
     to its thermal bath. This condition seems
     to be related to that considered in this paper
     where the dynamical system playing the role of
     bath is in fact so slow as to break the condition
     itself of time scale separation. The reaction of
     the scientific community, of which the authors of \cite{KW90} are a
     significant example, has been that the non canonical
     behavior detected numerically by Zhu and Robinson
     \cite{ZR89} is an artefact of numerical inaccuracy
     and limited computation time. This paper shows, on the contrary,
     that the opposite condition might apply, namely, that ordinary rather
     that anomalous statistics might be the result of numerical inaccuracy.
         We know that the round-off errors are equivalent
         to the influence of fluctuations of a given intensity
         $\epsilon$.
          The larger the computer accuracy, the  smaller is the
          intensity of the equivalent fluctuations. On the
          other hand, we know \cite{FMG95} that the effect of
          these fluctuations is that of changing the
          correlation function of Eq.(\ref{correlationfunction})
          into the correlation function $\Phi^{*}_{\xi}(t)$ related 
	  to the original by
          \begin{equation}
          \Phi^{*}_{\xi}(t) = \Phi_{\xi}(t) \exp(-t/t_{C}),
          \label{elenaeffect}
          \end{equation}
with $t_{C}$ proportional to $\epsilon^{\delta}$ and $\delta$
being a positive coefficient, of the order of unity, determined by
  the microscopic dynamics under study\cite{FMG95}.
   It is evident that at
  times $t > t_{C}$ the Markov approximation is valid,
   and as an effect of it the non standard equation
    of (\ref{fundamentalequation}) becomes identical to a conventional
    Fokker-Planck equation. The non conventional
    equilibrium of (\ref{exact}) is a time asymptotic property,
    and at any given time $t >> 1/\gamma$ we can produce a
    transition from the regime of non-ordinary
    statistics to a regime of canonical
    Gaussian equilibrium by increasing the
    intensity of the parameter $\epsilon$, so as to 
    realize the condition $1/\gamma > \tau_{C}$.

    Finally, we want to stress a problem worth of future
    investigation. This has to do with the  increasing
    attention devoted to the non-extensive thermodynamics of
    Tsallis\cite{tsallis1,tsallis2,tsallis3}. Non-extensive
    thermodynamics means that the deviation
    from the canonical equilibrium distribution is not more perceived
    as a violation of statistical mechanics. This is a very valuable
    aspect of  this research work\cite{tsallis1,tsallis2,tsallis3}.
    In fact, as
    a result of the interest that Tsallis's non-extensive
    statistical mechanics is raising, a deviation from the ordinary
    prescription, of the kind earlier mentioned, would be judged these
    days as a
  possible manifestation of non-extensive thermodynamics triggered by
 the long-range correlations of the dynamical system under study,
 rather than a consequence of numerical inaccuracy.

 However, the arguments of this paper show that under
 the specific form here adopted
 to establish a fluctuation-dissipation process
 in a case of dynamics without time scale,
 the basin of
 attraction for equilibrium distribution
 is given  by L\'{e}vy statistics.
 It is interesting to point out that L\'{e}vy statistics share with
 Tsallis statistics the power law behavior of the distribution tails.
 However, the central part of the L\'{e}vy distribution significantly
depart from
 the generalized canonical distribution of Tsallis.
 In an earlier paper\cite{ANNA} it has been shown that the adoption
 of Tsallis' non-extensive thermodynamics naturally leads, via
 entropy maximisation under a proper constraint, to a
 transition probability with an inverse power law decay at large
 distances. By repeated application of this kind of transition, as a
 consequence of the L\'{e}vy-Gnedenko theorem\cite{gnekol} the
 diffusion process is attracted by the basin of  L\'{e}vy statistics.
In the case of extremely
 weak friction, equilibrium is reached as a result
of a very large number of elementary transitions, and this
 is probably the main reason why eventually the resulting statistics
 is of  L\'{e}vy rather than Tsallis kind.

\end{document}